\documentclass[apj]{emulateapj}
\usepackage{apjfonts}
\usepackage{natbib}

\begin{document}

\def\mpchi{\,h^{-1}{\rm {Mpc}}}
\def\mpchii{\,h{\rm {Mpc}^{-1}}}
\def\kpchi{\,h^{-1}{\rm {kpc}}}
\def\kms{\,{\rm {km\, s^{-1}}}}
\def\msun{\,h^{-1}{\rm M_\sun}}
\def\k{\mathbf{k}}
\def\x{\mathbf{x}}
\def\q{\mathbf{q}}
\def\r{\mathbf{r}}
\def\s{\mathbf{s}}
\def\be{\begin{equation}}
\def\ee{\end{equation}}
\def\ba{\begin{eqnarray}}
\def\ea{\end{eqnarray}}

\def\sarc{$^{\prime\prime}\!\!.$}
\def\arcsec{$^{\prime\prime}$}
\def\arcmin{$^{\prime}$}
\def\degr{$^{\circ}$}
\def\seco{$^{\rm s}\!\!.$}
\def\ls{\lower 2pt \hbox{$\;\scriptscriptstyle \buildrel<\over\sim\;$}}
\def\gs{\lower 2pt \hbox{$\;\scriptscriptstyle \buildrel>\over\sim\;$}}
\def\mbh{$M_{\rm BH}$}
\def\mhalo{$M_{\rm halo}$}
\def\mhaloe{M_{\rm halo}}
\def\mctwo{$M_{\rm 200c}$}
\def\mctwoe{M_{\rm 200c}}
\def\mtwo{$M_{\rm 200}$}
\def\mtwoe{M_{\rm 200}}
\def\mstar{$M_{\rm star}$}
\def\mstare{M_{\rm star}}
\def\msune{M_{\odot}}
\def\msun{$M_{\odot}$}
\def\sis{$\sigma$}
\def\kms{$km\, s^{-1}$}
\def\vvir{$V_{\rm vir}$}
\def\vc{$V_c$}
\def\vce{V_c}
\def\vmax{$V_{\rm max}$}
\def\vmax{V_{\rm max}}
\def\msunhe{M_{\odot}/h}
\def\msunh{$M_{\odot}/h$}

\vspace{0.5cm}

\title{On the intermediate-redshift central stellar mass-halo mass relation, and implications for the evolution of the most massive galaxies since $z\sim 1$}

\author{Francesco Shankar\altaffilmark{1}, Hong Guo\altaffilmark{2}, Vincent Bouillot\altaffilmark{3}, Alessandro Rettura\altaffilmark{4,5}, Alan Meert\altaffilmark{6}, Stewart Buchan\altaffilmark{1}, Andrey Kravtsov\altaffilmark{7}, Mariangela Bernardi\altaffilmark{6}, Ravi Sheth\altaffilmark{6,8}, Vinu Vikram\altaffilmark{6}, Danilo Marchesini\altaffilmark{9}, Peter Behroozi\altaffilmark{10}, Zheng Zheng\altaffilmark{2}, Claudia Maraston\altaffilmark{11}, Bego\~{n}a Ascaso\altaffilmark{12}, Brian C. Lemaux\altaffilmark{13}, Diego Capozzi\altaffilmark{11}, Marc Huertas-Company\altaffilmark{12}, Roy R. Gal\altaffilmark{14}, Lori M. Lubin\altaffilmark{15}, Christopher J. Conselice\altaffilmark{16}, Marcella Carollo\altaffilmark{17}, Andrea Cattaneo\altaffilmark{13}}

\altaffiltext{1}{School of Physics and Astronomy, University of Southampton, Southampton SO17 1BJ, UK; F.Shankar@soton.ac.uk}
\altaffiltext{2}{Department of Physics and Astronomy, University of Utah, UT 84112, USA}
\altaffiltext{3}{Centre for Astrophysics, Cosmology \& Gravitation, Department of Mathematics \& Applied Mathematics,
University of Cape Town, Cape Town 7701, South Africa}
\altaffiltext{4}{Jet Propulsion Laboratory, California Institute of Technology,
MS 169-234, Pasadena, CA 91109, USA}
\altaffiltext{5}{Department of Astronomy, California Institute of Technology, MS 249-17, Pasadena, CA 91125, USA}
\altaffiltext{6}{Department of Physics and Astronomy, University of Pennsylvania, 209
South 33rd St, Philadelphia, PA 19104}
\altaffiltext{7}{Department of Astronomy \& Astrophysics, The University of Chicago,
Chicago, IL 60637 USA}
\altaffiltext{8}{International Center for Theoretical Physics, 34151 Trieste, Italy}
\altaffiltext{9}{Department of Physics and Astronomy, Tufts University, Medford, MA
02155, USA}
\altaffiltext{10}{Kavli Institute for Particle Astrophysics and Cosmology, Stanford, CA 94305, USA}
\altaffiltext{11}{Institute of Cosmology and Gravitation, Dennis Sciama Building, Burnaby Road, Portsmouth PO1 3FX, UK}
\altaffiltext{12}{GEPI, Observatoire de Paris, CNRS, Universit\'e Paris Diderot, 61, Avenue de l'Observatoire 75014, Paris  France}
\altaffiltext{13}{Aix Marseille Université, CNRS, LAM (Laboratoire d'Astrophysique de Marseille) UMR 7326, 13388, Marseille, France}
\altaffiltext{14}{Institute for Astronomy, University of Hawai'i, 2680 Woodlawn Drive, Honolulu, HI 96822, USA}
\altaffiltext{15}{University of California, One Shields Avenue, Davis, CA 95616, USA}
\altaffiltext{16}{University of Nottingham, School of Physics \& Astronomy, Nottingham, NG7 2RD UK}
\altaffiltext{17}{Institute for Astronomy, ETH Zurich, CH-8093 Zurich, Switzerland}

\begin{abstract}
The stellar mass-halo mass relation is a key constraint in all semi-analytic, numerical, and semi-empirical models of galaxy formation and evolution. However, its exact shape and redshift dependence remain debated. Several recent works support a relation in the local Universe steeper than previously thought. Based on the comparisons with a variety of data on massive central galaxies, we show that this steepening holds up to $z\sim 1$, for stellar masses $\mstare \gtrsim 2 \times 10^{11}\, \msune$. Specifically, we find significant evidence for a high-mass end slope of $\beta \gtrsim 0.35-0.70$, instead of the usual $\beta \lesssim 0.20-0.30$ reported by a number of previous results. When including the independent constraints from the recent BOSS clustering measurements, the data, independent of any systematic errors in stellar masses, tend to favor a model with a very small scatter ($\lesssim 0.15$ dex) in stellar mass at fixed halo mass, in the redshift range $z < 0.8$ and for $\mstare > 3 \times 10^{11}\, \msune$, suggesting a close connection between massive galaxies and host halos even at relatively recent epochs. We discuss the implications of our results with respect to the evolution of the most massive galaxies since $z \sim 1$.
\end{abstract}



\section{Introduction}
\label{sec|intro}

\begin{figure*}
    \center{\includegraphics[width=17truecm]{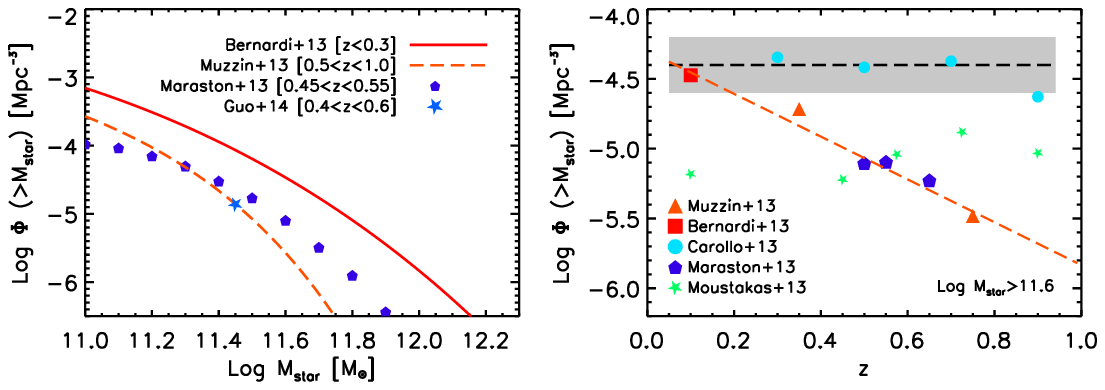}}
    \caption{\emph{Left}: Comparison among different cumulative stellar mass functions in the recent literature
    at different redshifts, as labelled. \emph{Right}: Redshift evolution of the cumulative number density of galaxies with $\mstare \gtrsim 4\times 10^{11}\, \msune$. The data by \citet{Bernardi13} (\emph{square}) and \citet{Carollo13} (\emph{circles}) would suggest at face value a negligible evolution in the cumulative number density, at least up to $z\lesssim 0.8$, but not so when considering other data. Where needed, we converted from a \citet{Kroupa01} to a \citet{Chabrier03} IMF via a constant shift of $-0.05$ dex \citep[e.g.,][]{Bernardi10}, and corrected for the (usually small) differences in cosmology.}
    \label{fig|NumberGalReds}
\end{figure*}

Probing the exact relation between stellar mass and host halo mass is one of the hottest topics in present-day cosmology \citep{Lea12,Yang12,Behroozi13,Moster13,Reddick13}. Such mapping can possibly shed light on the complex and still poorly understood physical processes that govern galaxy evolution \citep[e.g.,][]{SilkReviewGalx}, as well as unveil key properties of the underlying dark matter cosmological model \citep[e.g.,][]{Weinberg13}.

Constraining the statistical and environmental evolution of massive galaxies, especially those of $\mstare \gtrsim (2-3)\times 10^{11}\, \msune$, is particularly meaningful. A number of independent observations are showing that galaxies above this mass scale tend to depart from simple extrapolations of the scaling relations characterizing their lower-mass counterparts, having larger sizes, more prolate shapes, and redder colors \citep[e.g.,][]{Van09,Bernardi11a}.

However, the galaxy-halo mapping for massive galaxies as inferred from abundance matching between the stellar and halo mass functions, is still under debate. One of the main uncertainties relies on a proper determination of the stellar mass function \citep[e.g.,][]{Bernardi13,Muzzin13}. For example, the constant number density evolution of the massive galaxies derived by, e.g., \citet{Carollo13} at $z \lesssim 1$, is in disagreement with other measurements at similar redshifts \citep{Maraston13,Muzzin13}.

In this letter, we provide additional, key constraints to the \mstar-\mhalo\ relation for massive central galaxies at $0<z<1$ using direct stellar and host halo mass measurements of the Brightest Cluster Galaxies (BCGs), as well as accurate galaxy clustering measurements at $0.4<z<0.8$. The galaxy clustering measurements are used to infer the host halo mass distributions through the halo occupation distribution (HOD) models \citep[e.g.,][]{Zheng07}, and thus provide a powerful tool to break the degeneracies inherent to the abundance matching techniques.

In the following we will adopt a cosmology with parameters
$\Omega_{\rm m}=0.30$, $\Omega_{\rm b}=0.045$, $h=0.70$,
$\Omega_\Lambda=0.70$, $n_s=1$, and $\sigma_8=0.8$, to match the one assumed in our reference
stellar mass functions and halo occupation measurements.
We will adopt the Chabrier Initial Mass Function (IMF; \citealt{Chabrier03}) as our reference one.

\section{Method}
\label{sec|Method}

To provide constraints on the galaxy-halo mapping at $z>0$, we evaluate the median stellar mass at fixed host halo mass by direct abundance matching between the stellar
and halo mass functions at a given redshift
\begin{equation}
\Phi(>\mstare,z)=\Phi_c(>\mctwoe,z)+\Phi_s(>\mctwoe,z)
\label{eq|Cum}
\end{equation}
with $\mctwoe$ the halo masses defined as 200 times the critical density at redshift $z$.
The $\Phi_c(>\mctwoe,z)$ term refers to the host halo mass function, which we take from \citet{Tinker08}, as it can be adapted to diverse
halo definitions, and it is well defined up to $\mctwoe \lesssim 10^{15}\, \msune$.
Eq.~\ref{eq|Cum} includes the subhalo term $\Phi_s(>\mctwoe,z)$ with unstripped mass $\mctwoe$, which we take from \citet{Behroozi13}.
Neglecting the satellite term in Eq.~\ref{eq|Cum} makes very little difference in the halo mass range of interest here, e.g., $\mctwoe \gtrsim 10^{13}\, \msune$.

It is instead much more relevant to adopt the proper intrinsic scatter $\Sigma$ in stellar mass at fixed halo mass, ideally constrained from independent datasets, as larger values of $\Sigma$ induce a flatter $\mstare-\mhaloe$ relation above the break.
Eq.~\ref{eq|Cum} does not assume any scatter between stellar and halo mass, however
one straightforward way to include it is as follows (see also, e.g., \citealt[][]{Behroozi10}).
At any redshift of interest, we first fit the parameters of a two-power law relation defined as
\begin{equation}
\mstare=\mstare^0 \left(\frac{\mctwoe}{\mctwoe^0}\right)^{\alpha} \left[1+\left(\frac{\mctwoe}{\mctwoe^0}\right)^{\gamma}\right]^{-1}
    \label{eq|MstarMhalo}
\end{equation}
to the raw output of Eq.~\ref{eq|Cum}.
We then choose a value for the intrinsic scatter $\Sigma$, and generate a large galaxy catalog by assigning to each
(sub)halo extracted from the total halo mass function, a galaxy with stellar mass derived from
a Gaussian distribution with mean given by the logarithm of Eq.~\ref{eq|MstarMhalo}, and dispersion $\Sigma$ (in dex).
We finally vary $\gamma$ in Eq.~\ref{eq|MstarMhalo} to tune the high mass-end slope $\beta=\alpha-\gamma$ until the input stellar mass function in Eq.~\ref{eq|Cum} is fully reproduced.

\section{Data}
\label{sec|Data}

The data on BCGs in groups and clusters considered in this letter are derived at $z=0.1$ from X-rays \citep{Kravtsov14}, at $0.2<z<1$ from X-ray and weak lensing in COSMOS \citep{Finoguenov07,George11,Huertas13a}, at $0.8<z<1.4$ from IR \citep[SpARCS;][]{Lidman12,Burg13} and X-ray data \citep{Strazzu10,Raichoor11,Rettura11}, and at $z\sim1$ from the Cl1604 supercluster and other structures from the ORELSE survey \citep{Ascaso13}.

As for clustering, we utilize the massive galaxies at the median redshift of $z\sim0.6$ from the CMASS sample of the
Sloan Digital Sky Survey-III (SDSS-III) Baryon Oscillation Spectroscopic Survey \citep[BOSS;][]{Dawson13}. Stellar masses are from the Portsmouth SED-fitting \citep{Maraston13},
originally derived assuming a Kroupa IMF \citep{Kroupa01}.
The host halo masses for these massive galaxies are estimated through the
HOD modeling of the projected-space two-point
correlation functions on scales from $0.1\mpchi$ to $60\mpchi$,
faithfully following the method laid out in \cite{GuoHong14}.

\begin{figure*}
    \center{\includegraphics[width=17truecm]{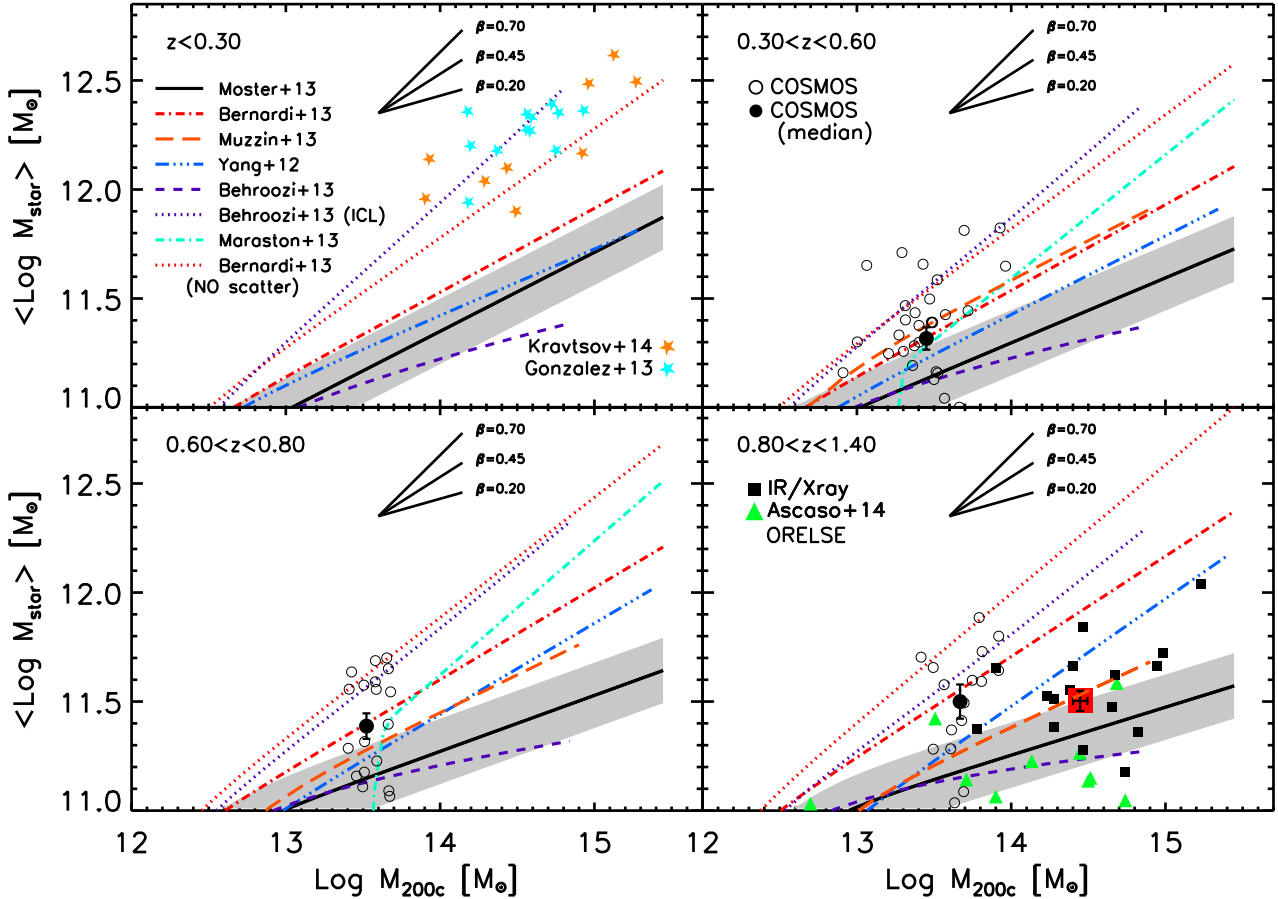}}
    \caption{Median stellar mass as a function of halo mass relation for \emph{central} galaxies at $z=0.1,0.4,0.7,1.1$, clockwise from the \emph{upper left} panel, respectively. The \emph{red}, \emph{dot-dashed} and \emph{red}, \emph{dotted} lines are derived from the \citet{Bernardi13} stellar mass function, assumed constant at all redshifts above $\mstare \gtrsim 10^{11}\, \msune$, with an intrinsic scatter of $\Sigma=0.25$ and $\Sigma=0$ dex, respectively. The \emph{solid} line and \emph{gray area} mark the \citet{Moster13} median relation and its $1-\sigma$ uncertainty region, respectively. The \emph{blue}, \emph{dashed} line is the result by \citet{Yang12}. All the data are as labeled. The \emph{filled circles} mark the median stellar mass-halo mass in the COSMOS data, while the \emph{red square} the median in the IR/X-ray plus \citet{Ascaso13} and other associated data.}
    \label{fig|MsMhaloRel}
\end{figure*}

\section{Results}
\label{sec|results}

\subsection{The number density of massive galaxies}
\label{subsec|NgalMassive}

The first step towards defining a more secure mapping between stars and halos relies on properly measuring the stellar mass function of galaxies.
The left panel of Fig.~\ref{fig|NumberGalReds} shows the cumulative number density of galaxies from \citet{Bernardi13} for the SDSS-DR7 main
galaxy sample ($z \lesssim 0.2$; solid, red line). We used their estimate based on S\'{e}rsic-exponential light profile, which is considered by the authors to be the most realistic one to describe SDSS data \citep{Bernardi12}.
When compared to the COSMOS/UltraVISTA data by \citet{Muzzin13} (long-dashed line), at the average redshift of $z=0.75$, or the BOSS estimate from \citet{GuoHong14} (star), or even the BOSS determination of the stellar mass function by \citet{Maraston13} (diamonds), it would imply at face value a significant increase in the number
density of massive galaxies towards low redshifts.

The right panel of Fig.~\ref{fig|NumberGalReds} focuses on the
number density evolution of galaxies above $\mstare \gtrsim 4\times 10^{11}\, \msune$.
For completeness, this panel also reports the measurements inferred by \citet{Mou13} and \citet{Carollo13}, which would instead suggest a negligible evolution since $z \lesssim 1$. \citet{Mou13} is well consistent with the stellar mass function by \citet{Maraston13}. \citet{Carollo13} do not subtract stellar mass losses from the total masses, thus explaining at least part of the inconsistency with other determinations.
%
%

Overall, the right panel of Fig.~\ref{fig|NumberGalReds} brackets the possible evolutionary paths since $z \lesssim 1.0$ for the number density of massive galaxies, from a non evolving scenario (black, long-dashed line), to a fast evolving one (orange, dashed line). Most relevant measurements broadly fall within these sequences \citep[e.g.,][]{Ilbert13}. The exact determination of the evolution and normalization of the high-mass end of the stellar mass function is limited by photometric and spectral systematics in the determination of stellar masses, as well as possible incompleteness and/or cosmic variance issues \citep[e.g.,][]{Marchesini09,Bernardi10,Bernardi13,Ilbert13,Muzzin13}. In the following, we will evaluate the stellar mass-halo mass relation considering both of these extreme cases, and, by direct comparison with independent data sets, namely large scale clustering, set constraints on plausible evolutionary paths for the most massive galaxies in light of current estimates of the stellar mass function.

\subsection{The stellar mass-halo mass relation}
\label{subsec|MstarMhaloRel}

Fig.~\ref{fig|MsMhaloRel} shows the median stellar mass as a function of host halo mass relation for central galaxies evaluated at $z=0.1$ (upper left), $z=0.4$ (upper right), $z=0.7$ (lower left), $z=1.1$ (lower right), for different models. 
The dot-dashed, red lines are obtained by inserting in Eq.~\ref{eq|Cum} the \citet{Bernardi13} stellar mass function, assumed to be constant up to $z\sim1$,
and inclusive of a scatter of $\Sigma=0.25$ dex in stellar mass at fixed halo mass. The dotted lines refer to the same model but without scatter.

The long-dashed, orange lines adopt instead the \citet{Muzzin13} stellar mass function, only valid at $z>0.2$, with an intrinsic scatter of $\Sigma=0.15$ dex. For completeness, we compare these results with three mappings from the recent literature, the \citet{Moster13} median relation (solid, black lines), with its $1~\sigma$ error bar (gray area), the \citet{Yang12} relation (dot-dashed, blue lines), and the \citet{Behroozi13} model (dashed, purple lines). Other recent works mostly lie within the \citet{Moster13} uncertainty region \citep[e.g.,][]{Lea12}. For completeness, we also show with purple dotted lines, the \citet{Behroozi13} model inclusive of the total intra-cluster light.

\begin{figure*}
    \center{\includegraphics[width=17truecm]{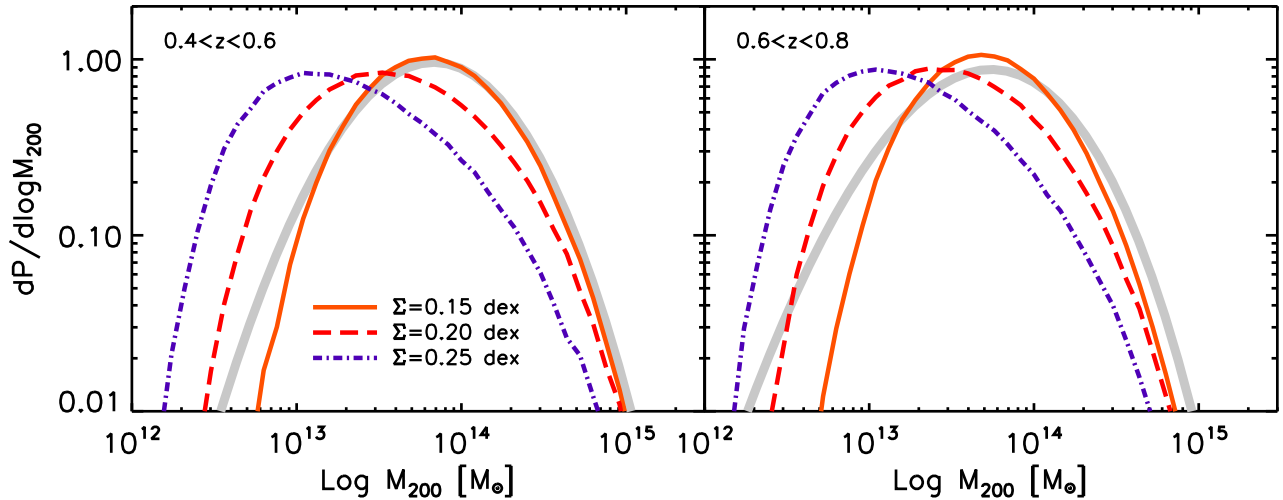}}
    \caption{Predicted halo mass distributions from the three mapping of interests in this paper, as labeled, compared to the BOSS clustering data from \citet{GuoHong14} at $z=0.5$ (\emph{left}) and $z=0.7$ (\emph{right}). Models with too large intrinsic scatter in stellar mass at fixed halo mass are challenged by the BOSS data.}
    \vspace{0.4cm}
    \label{fig|BOSS}
\end{figure*}

Overall, most of the recent estimates of the stellar-halo mass relation tend to be discrepant with respect to direct central galaxy mass measurements in groups and clusters. All the available data collected in this work in fact, although with a large dispersion, tend to lie, on average, above the \citet{Moster13} uncertainty region, implying a steeper stellar-halo mass relation, with the high-mass end slope (Eq.~\ref{eq|MstarMhalo}) increasing from $\beta \lesssim 0.2-0.3$ to $\beta \gtrsim 0.35-0.70$. Such a discrepancy was already emphasized at $z<0.3$ by some groups \citep[e.g.,][]{Kravtsov14,Shankar14}. \citet{Kravtsov14}, in particular, recomputed abundance matching with the \citet{Bernardi13} stellar mass function, finding a steeper relation above $\mstare \gtrsim 10^{11}\, \msune$, broadly consistent with their direct nine BCG stellar and halo mass measurements (orange stars). Our own determinations of the stellar-halo mass relation via Eq.~\ref{eq|Cum} based on the local \citet{Bernardi13} stellar mass function without scatter (dotted, red lines in Fig.~\ref{fig|MsMhaloRel}), are at $z=0.1$ broadly consistent with the \citet{Kravtsov14} and \citet{Gonzalez13} data at very high masses.

One of the primary cause of the discrepancies can be ascribed
to the adoption of different input stellar mass functions. In particular, the \citet{Bernardi13} stellar mass function, based on improved sky subtractions and modeling of the central galaxy light profile, is characterized by a significant boost in the abundance of the most massive galaxies, which in turn induces a steepening of the stellar mass-halo mass relation.
Other factors contribute to the differences in Fig.~\ref{fig|MsMhaloRel}. \citet{Moster13}, for example, took care in de-convolving their adopted stellar mass function by some systematic errors before applying Eq.~\ref{eq|Cum}, thus producing a flattening in the high-mass end of their inferred stellar mass-halo mass relation.

\subsection{Independent constraints from clustering}
\label{subsec|cluste}

Fig.~\ref{fig|MsMhaloRel} also reveals that at $z \sim 0.3-0.6$, a clear degeneracy exists between a model based on \citet{Muzzin13}, with an intrinsic scatter in stellar mass at fixed halo mass of $\Sigma=0.15$ dex (long-dashed, orange lines), and the one based on the $z=0.1$ \citet{Bernardi13} stellar mass function with $\Sigma=0.25$ dex (dot-dashed, red lines). In fact, both models can potentially reproduce the COSMOS data,
though the latter with larger scatter would imply a constant number density at least up to $z\sim 0.8$, at variance with the former. Irrespective of uncertainties on stellar masses, we discuss in this section how to use clustering to set a secure upper limit to $\Sigma$.

Fig.~\ref{fig|BOSS} displays with gray bands the \citet{GuoHong14} HOD host halo mass distributions for \emph{central} galaxies\footnote{Given the numerous complexities and variables at play in properly modeling satellites in abundance matching, e.g., redshift of infall, effect of environment, etc... \citep[e.g.,][]{Neistein11,Yang12}, we here discuss predictions for only central galaxies, and focus on the large-scale clustering and bias. The fraction of satellites in our stellar mass range is anyway very small \citep{GuoHong14}.} with stellar mass above $\log \mstare > 11.50$ (Kroupa IMF) at $z=0.5$ (left) and $z=0.7$ (right) inferred from the BOSS CMASS clustering measurements (Sect.~\ref{sec|Data}). For the stellar mass of interest here, the galaxy sample is almost complete and the tiny fraction of missing galaxies due to the CMASS sample selections have negligible effects on the clustering measurements \citep{Maraston13,GuoHong14}. We compare the BOSS results with the abundance matching model based on the \citet{Muzzin13} stellar mass function, which perfectly matches the cumulative number density adopted by \citet{GuoHong14} (left panel of Fig.~\ref{fig|NumberGalReds}). At each redshift of interest we generate a mock halo
catalog extracted from the halo mass function, and populate the halos with galaxies through the stellar mass-halo mass relation based on \citet{Muzzin13} with a given dispersion $\Sigma$. 

Our results are shown in Fig.~\ref{fig|BOSS} for three different values of the scatter $\Sigma=0.15,0.20,0.25$ dex, as labeled. Consistently with the reference HOD model, all our mock catalogs have halo masses defined as 200 times the background density at the redshift of interest, and matched to the stellar mass cut in BOSS.
Models based on scatters larger than $\Sigma>0.15$ dex, inevitably map galaxies at fixed stellar mass to host halo masses significantly lower than that inferred from clustering measurements. A larger scatter tends to overall \emph{flatten} the $\mstare-\mhaloe$ relation above the break. However, increasing the scatter also includes lower-mass, more numerous halos in samples defined by stellar mass thresholds, thus effectively \emph{lowering} the median halo mass at fixed stellar mass.

The lower scatter of $\Sigma=0.15$ dex is fully consistent with the inferred scatter $\sigma_{\log M_{200}}$ (${\sim}0.62$ at $z=0.5$ and ${\sim}0.76$ at $z=0.7$) in the HOD model, which describes the scatter in the host halo mass distribution for the stellar mass sample. The scatter $\sigma_{\log M_{200}}$ can be converted into $\Sigma$ through $\Sigma=p\sigma_{\log M_{200}}/\sqrt{2} \sim 0.17$ when assuming a power-law relation of $\mstare\propto M_{200}^p$ \citep{Zheng07}, with $p \sim 0.35$ as found for \citet[][cfr. Fig.~\ref{fig|MsMhaloRel}]{Muzzin13}.
Our results of a low scatter are in line with and extend several previous estimates \citep[e.g.,][]{More09,Moster10,Lea12,Yang12,Rodriguez14}.

\begin{figure}
    \center{\includegraphics[width=8.5truecm]{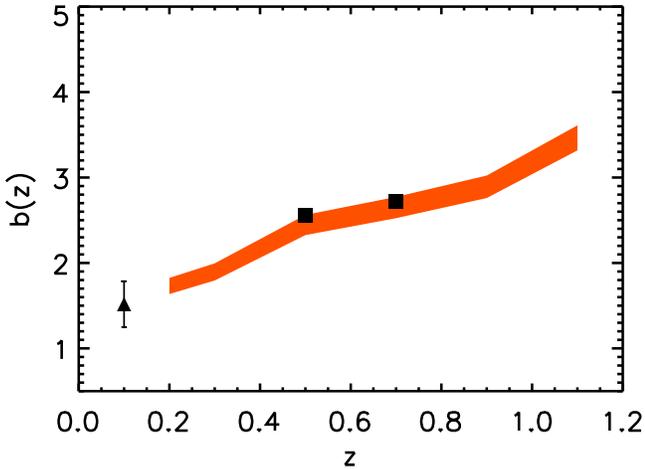}}
    \caption{Median bias as a function of redshift for galaxies above $\mstare \gtrsim 3 \times 10^{11}\, \msune$, as predicted by our reference stellar mass-halo mass mapping.
    The \emph{colored contour} defines the systematic uncertainty associated to the \citet{Sheth01} and \citet{Tinker05} biases.
    Data are extracted from the results by \citet{Yang07} (\emph{filled triangle}), and \citet{GuoHong14} (\emph{filled squares}).}
    \label{fig|biasz}
\end{figure}

Analogously, the low-scatter model is fully consistent with the predicted large scale bias as a function of redshift (Fig.~\ref{fig|biasz})
derived by \citet{GuoHong14} from BOSS data (filled squares), all defined for galaxies
above $\mstare \gtrsim 3 \times 10^{11}\, \msune$. The colored contour defines the systematic uncertainty associated to the \citet{Sheth01} and \citet{Tinker05} biases.
For completeness, in the same Fig.~\ref{fig|biasz} we also report the $z \sim 0.1$ bias (filled triangle) extracted from the \citet{Yang07} catalogue, and matched to the \citet{Bernardi12} SDSS revised stellar masses (see also \citealt{Huertas13b} and \citealt{Shankar14}).

Our result on a low scatter in the stellar-halo mass relation is independent of systematics in stellar masses, at least for \emph{central} galaxies. In fact, any error in stellar mass will equally propagate in the cumulative number density and connected HOD clustering modeling. Higher stellar masses, for example, will induce larger number densities and proportionally lower, large-scale characteristic correlation lengths (thus lower median host halo masses) above a fixed limit in stellar mass \citep{Bernardi13}. Our stellar-halo median relation, based on Eq.~\ref{eq|MstarMhalo}, will also map galaxies to lower host halo masses, but will still require a low scatter $\Sigma$ to fully match the HOD results.


\section{Discussion}
\label{sec|discu}

In this letter we found significant evidence for:
\begin{itemize}
\item a steeper stellar mass-halo mass relation with $\beta \gtrsim 0.35-0.70$ instead of $\beta \lesssim 0.2-0.3$ from previous works;
\item a low scatter $\Sigma \lesssim 0.15$ dex in stellar mass at fixed host halo mass, at least up to $z \sim 0.8$.
\end{itemize}
Our results can potentially set valuable constraints to the viable evolutionary paths of massive galaxies.


We first take the \citet{Bernardi13} stellar mass function as the $z \sim 0$ reference, as it well matches all local data on massive BCGs (upper left panel of Fig.~\ref{fig|MsMhaloRel}). A steadily decreasing number density of massive galaxies at $0.3<z<0.8$ \citep[e.g.,][right panel of Fig.~\ref{fig|NumberGalReds}]{Muzzin13}, would then, at face value, be consistent with most of the available constraints on the group and cluster centrals, keeping $\Sigma \lesssim 0.15$ dex to match the HOD halo mass distributions inferred from the BOSS clustering measurements (Fig.~\ref{fig|BOSS}).

Another extreme case is forcing the \citet{Bernardi13} number density of massive galaxies to be constant up to $z \sim 1$ \citep[e.g.,][right panel of Fig.~\ref{fig|NumberGalReds}]{Carollo13}. However, the latter model, coupled to the need for a negligible scatter $\Sigma$, would imply a systematic overestimate of a factor of $\gtrsim 5$ in the median BCG stellar mass, as currently measured in clusters at $z \gtrsim 0.8$ for $\log M_{200c}/M_\odot \gtrsim 14.5$ (red dotted line versus red square in the bottom right panel of Fig.~\ref{fig|MsMhaloRel}), and an overestimate of a factor $\sim 2$ of the total stellar plus intra-cluster light model by Behroozi et al. (2013; purple dotted lines in Fig.~\ref{fig|MsMhaloRel}).

Irrespective of the systematics in the stellar mass function, \emph{current} BCG mass determinations and HOD clustering measurements, may favor an increase of a factor of a few since $z \lesssim 1$ in the number density of the most massive galaxies. This can be partly induced by a parallel growth in the median stellar mass. Independent semi-empirical studies indeed suggest an increase in stellar mass by a factor of $\sim 2$ since $z \lesssim 1$ \citep[e.g.,]{Zheng07,Lidman13,Ascaso13,Marchesini14}. As supported by state-of-the-art hierarchical galaxy evolution models \citep[e.g.,][]{DeLucia06,Shankar13}, a non-negligible contribution to this mass growth can be explained by minor and major mergers. The latter, in particular, might be the ones responsible for the steepening in the high mass-end of the scaling relations characterizing early-type galaxies \citep[e.g.,][]{Bernardi11b}.

\vspace{0.5cm}
\begin{acknowledgements}
FS acknowledges Naresh Shankar, Jeremy Tinker, David Weinberg, Federico Marulli, and Surhud More for several interesting and helpful discussions. VB is supported financially by the National Research Foundation of South Africa. DM acknowledges the support of the Research Corporation for Science Advancement's Cottrell Scholarship. This work is based on data obtained with the {\it Spitzer Space Telescope}, which is operated by the Jet Propulsion Lab (JPL), California Institute of Technology (Caltech), under a contract with NASA. We thank the referee for a constructive report that significantly improved the presentation of the results.
\end{acknowledgements}

\bibliographystyle{yahapj}

\label{lastpage}
\end{document}